\documentclass[prb,twocolumn,preprintnumbers,amsmath,amssymb,floats,nobalancelastpage]{revtex4}
\usepackage{graphicx}% Include figure files
\usepackage{dcolumn}% Align table columns on decimal point
\usepackage{bm}
\usepackage{times}

\begin{document}
%\preprint{for PRB}

\title{Line nodes in the energy gap of high-temperature superconducting BaFe$_2$(As$_{1-x}$P$_x$)$_2$ from
penetration depth and thermal conductivity measurements}

\author{K.~Hashimoto$^{1,2,\ast}$}
\author{M.~Yamashita$^{1,\ast}$}
\author{S.~Kasahara$^{3}$} %,4}$}
\author{Y.~Senshu$^1$}
\author{N.~Nakata$^1$}
\author{S.~Tonegawa$^1$}
\author{K.~Ikada$^1$}
\author{A.~Serafin$^2$}
\author{A.~Carrington$^2$}
\author{T.~Terashima$^3$}
\author{H.~Ikeda$^1$}
\author{T.~Shibauchi$^1$}
\author{Y.~Matsuda$^1$}

\affiliation{$^1$Department of Physics, Kyoto University, Sakyo-ku, Kyoto 606-8502, Japan \\
$^2$H. H. Wills Physics Laboratory, University of Bristol, Tyndall Avenue, Bristol, UK \\
$^3$Research Center for Low Temperature and Materials Sciences, Kyoto University, Sakyo-ku, Kyoto 606-8501, Japan }

\date{\today}

%\wideabs{

%}

\begin{abstract}
We report magnetic penetration depth and thermal conductivity data for high-quality single crystals of
BaFe$_2$(As$_{1-x}$P$_{x}$)$_2$ ($T_c=30$\,K) which provide strong evidence that this material has line nodes in its
energy gap. This is distinctly different from the nodeless gap found for (Ba,K)Fe$_2$As$_2$ which has similar $T_c$ and
phase diagram. Our results indicate that repulsive electronic interactions play an essential role for Fe-based
high-$T_c$ superconductivity but that uniquely there are distinctly different pairing states, with and without nodes,
which have comparable $T_c$.
\end{abstract}

\maketitle

%%%%%%%%%Introduction

The most important question concerning the Fe-based high temperature superconductors \cite{Ishida09} is what is the
interaction that glues the electrons into Cooper pairs. Conventional phonon-mediated pairing leads to the
superconducting gap opening all over the Fermi surface, while unconventional pairing mechanisms, such as spin
fluctuations, can lead to a gap which has opposite signs on some regions of the Fermi surface. The sign change is a
result of the anisotropic pairing interaction which is repulsive in some momentum directions. In high-$T_c$ cuprate
superconductors, where the electronic structure is essentially described by a single quasi-two-dimensional Fermi
surface, the sign change of order parameter inevitably produces line nodes in the gap function. In Fe-pnictides,
however, the Fermi surface has disconnected hole and electron sheets and so the condition for a sign changing gap can
be fulfilled without nodes. In fact, a simple picture based on spin fluctuations predicts a distinct type of
unconventional order parameter with sign change between the sheets, known as an $s_\pm$ state.\cite{Mazin08,Kuroki09}
In this case, each Fermi surface is fully gapped, preventing low-energy excitation of quasiparticles.

Many experimental studies of high-$T_c$ Fe-arsenides indicate a fully-gapped superconducting
state.\cite{Ishida09,Hashimoto09a,Malone09,Ding08,Hashimoto09b,Luo09,Yashima09} However, some measurements have
suggested the existence of low-lying quasiparticle excitations \cite{Ishida09,Gordon09,Martin09} which is consistent
with a strongly disordered fully gapped $s_\pm$ or a nodal state.\cite{Hashimoto09b,Vorontsov09,Bang09}. Strong
evidence for gap-nodes in the clean limit has been reported for LaFePO.\cite{Fletcher09,Hicks09,Yamashita09} However,
this material has a low $T_c$ and has no nearby magnetic phases and so it is unclear whether it is representative of
the higher $T_c$ Fe-based superconductors.

Recently, high-quality single crystals of the isovalent pnictogen substituted system BaFe$_2$(As$_{1-x}$P$_x$)$_2$ with
$T_c$ as high as 30\,K have been grown.\cite{Kasahara09} As with the electron and hole doped materials,
superconductivity with similar maximum $T_c$ appears in close proximity to the spin-density-wave phase boundary
($x\approx 0.3$), where the presence of the strong antiferromagnetic fluctuations has been detected by NMR
measurements.\cite{Yashima09,Nakai09} It has been suggested from the spin-fluctuations theory \cite{Kuroki09} that the
superconducting gap structure can depend sensitively on the pnictogen height from the Fe plane. The substitution of P
for As reduces this pnictogen height,\cite{Kasahara09} which is not the case for the hole doping by K substitution for
Ba. A comparison of the gap structure in BaFe$_2$(As$_{1-x}$P$_x$)$_2$ and (Ba,K)Fe$_2$As$_2$ with similar $T_c$ should
then give important insight into the pairing mechanism of high-$T_c$ superconductivity in Fe-pnictides.

Here we report low-temperature measurements of the magnetic penetration depth $\lambda$ and thermal conductivity
$\kappa$ in high-quality crystals of BaFe$_2$(As$_{1-x}$P$_x$)$_2$ ($x=0.33$) with optimum $T_c=30$\,K. Both $\lambda$
and $\kappa$ are particularly good probes of the gap structure of superconductors. $\lambda$ is related to the
superfluid density $n_s\propto\lambda^{-2}$, whose temperature dependence is directly determined by the gap function.
$\kappa$ probes low-energy delocalized quasiparticles carrying entropy, which extend over the whole crystal. Both
measurements probe the bulk superconducting properties. Our results provide strong evidence for line nodes in the
energy gap in this system. The presence of nodes is in sharp contrast to the fully gapped superconducting state deduced
from similar measurements of (Ba,K)Fe$_2$As$_2$ having similar $T_c$, and points towards a electronic repulsive pairing
interaction in high-$T_c$ Fe-pnictide superconductors.

%%%%%%%%%%%%%%%%%%%%%%FIG 1%%%%%%%%%%%%%%%
\begin{figure}[tb]
%h=here, t=top, b=bottom, p=separate figure pag
\begin{center}%\leavevmode
\includegraphics[width=0.97\linewidth]{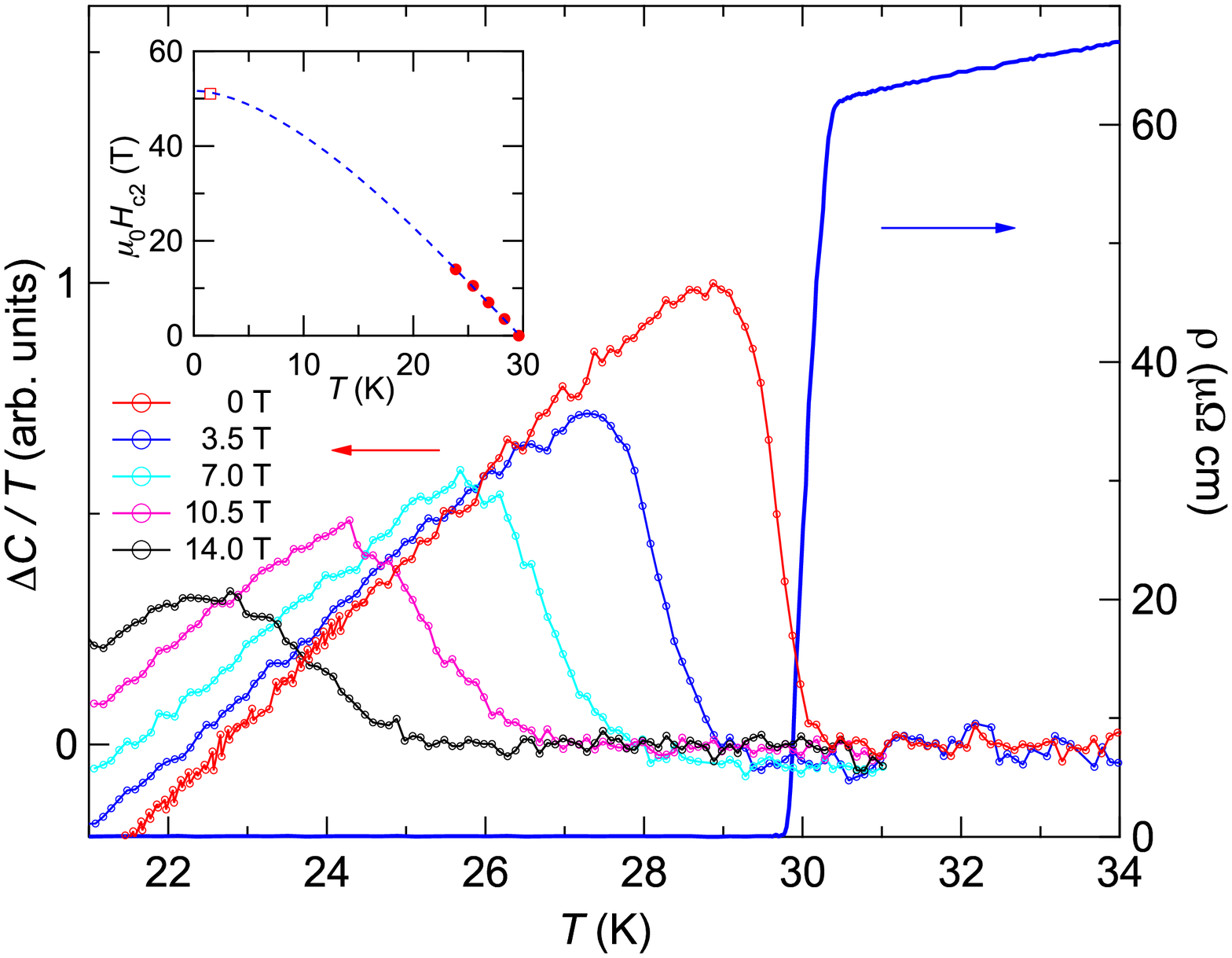}
\caption{ (Color online) Relative change in the electronic specific heat $\Delta C/T$ of single crystal
BaFe$_2$(As$_{0.67}$P$_{0.33}$)$_2$ measured using an ac technique under magnetic fields applied along the $c$ axis. A
smooth background determined from an extrapolation of the 14\,T data has been subtracted. Temperature dependence of the
in-plane resistivity $\rho$ of a crystal from the same batch is also shown. Inset shows the upper critical field
$H_{c2}(T)$ determined from the mid point of the specific heat jump at each field (closed circles) and the
irreversibility field measured by torque measurements (open square).\cite{Shishido09} The dashed line is a fit to the
Wertharmer-Helfand-Hohenberg formula. } \label{C_T}
\end{center}
\end{figure}
%%%%%%%%%%%%%%%%%%%%%%FIG 1%%%%%%%%%%%%%%%

\begin{figure}[tb]
%h=here, t=top, b=bottom, p=separate figure pag
\begin{center}%\leavevmode
\includegraphics[width=0.97\linewidth]{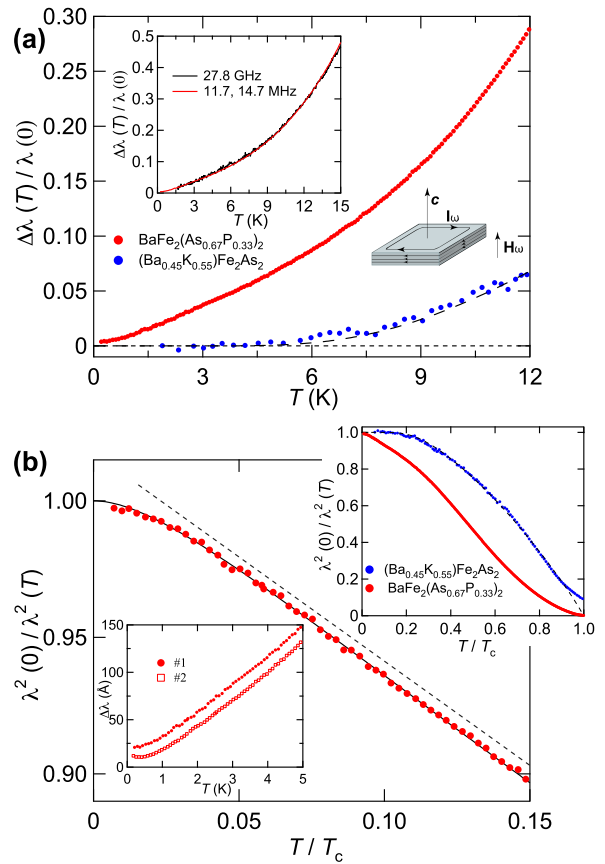}
\caption{ (Color online) (a) The change $\Delta\lambda(T)\equiv \lambda(T)-\lambda(0)$ in the penetration depth of
BaFe$_2$(As$_{0.67}$P$_{0.33}$)$_2$ along with the data for (Ba$_{0.45}$K$_{0.55}$)Fe$_2$As$_2$
($T_c\simeq33$\,K).\cite{Hashimoto09b} Inset compares the results obtained by the microwave and RF techniques. (b)
Normalized superfluid density $\lambda^2(0)/\lambda^2(T)$ as a function of $T/T_c$. The data follow a $T$-linear
dependence (dashed line) down to $T/T_c\sim 0.05$. The deviation at lower $T/T_c$ can be fitted to the dependence for a
gap with line nodes with disorder (solid line). The upper inset is the data up to $T_c$. The tail near $T_c$ in
(Ba$_{0.45}$K$_{0.55}$)Fe$_2$As$_2$ is due to the skin depth effect at microwave frequencies.\cite{Hashimoto09b} The
lower inset shows $\Delta\lambda(T)$ at low temperatures in two samples (shifted vertically for clarity). }
\label{lambda}

\end{center}
\end{figure}

Our BaFe$_2$(As$_{0.67}$P$_{0.33}$)$_2$ crystals were grown using a self flux method \cite{Kasahara09} and were
characterized using x-ray diffraction and energy dispersion (EDX). No impurity phases were detected within experimental
limits of $\lesssim 1$\%.  The samples exhibit excellent sharp bulk superconducting transitions [see Fig.\:\ref{C_T}].
at $T_c= 30$\,K in both the dc resistivity $\rho$ as well as the specific heat. Importantly for studies of the gap
structure, these samples are relatively free from disorder, as demonstrated by the observation of quantum oscillations
 \cite{Shishido09}. The temperature dependence of penetration depth $\lambda(T)$ in
BaFe$_2$(As$_{0.67}$P$_{0.33}$)$_2$ was measured by a MHz tunnel-diode oscillator down to $\sim 0.15$\,K
(Ref.\:\onlinecite{Fletcher09}) and by a microwave superconducting cavity resonator down to $\sim
1.6$\,K.\cite{Hashimoto09a,Hashimoto09b} In both measurements a weak ac magnetic field is applied along the $c$ axis,
generating supercurrents in the $ab$ plane. In the microwave measurements with angular frequency $\omega$, the absolute
values of both the real ($R_s$) and imaginary ($X_s$) parts of surface impedance can be determined by using the
relation $R_s=X_s=\sqrt{\mu_0 \omega \rho / 2}$.\cite{Hashimoto09a,Hashimoto09b} This allows us to estimate
$\lambda(0)=X_s(0)/\mu_0 \omega = 200\pm 30$\,nm, which is consistent with the recent $\mu$SR results of
$\lambda(0)\approx 170$\,nm.\cite{Bernhard} The $\lambda(T)$ results obtained by the two techniques at different
frequencies show excellent agreement [inset of Fig.\:\ref{lambda}(a)]. Thermal conductivity was measured in a dilution
refrigerator with the heat current applied in the $ab$ plane.\cite{Yamashita09}

%%%%%%%%%%%%Penetration depth

%%%%%%%%%%%%%%%%%%%%%%FIG 2%%%%%%%%%%%%%%%
%%%%%%%%%%%%%%%%%%%%%%FIG 2%%%%%%%%%%%%%%%

Figure\:\ref{lambda}(a) shows the normalized change in the penetration depth $\Delta\lambda(T)/\lambda(0)$ in
BaFe$_2$(As$_{0.67}$P$_{0.33}$)$_2$, compared with previous results for a clean (Ba$_{0.45}$K$_{0.55}$)Fe$_2$As$_2$
crystal.\cite{Hashimoto09b} In sharp contrast to the flat behavior observed in the K-doped crystal, $\Delta\lambda(T)$
in the P-substituted crystal exhibits a strong quasi-linear temperature dependence at low temperatures. The $T$-linear
dependence of $\Delta\lambda(T)$ is a strong indication of line nodes in the superconducting gap. The normalized
superfluid density $\lambda^2(0)/\lambda^2(T)$ in Fig.\:\ref{lambda}(b) also clearly demonstrates the fundamental
difference between P- and K-doped samples. The low-temperature data of BaFe$_2$(As$_{0.67}$P$_{0.33}$)$_2$ can be
fitted to $1-\alpha (T/T_c)^n$ with the exponent $n=1.13(\pm 0.05)$ close to unity. This is completely incompatible
with the flat exponential dependence observed in the fully gapped superconductors, and immediately indicates low-lying
quasiparticle excitations in this system. This behavior is fundamentally different from the power-law dependence of
superfluid density with powers varying $n\sim2.0$ to $\sim2.4$ found for other Fe-arsenides.\cite{Hashimoto09b,Gordon09,Martin09} In the fully gapped unconventional $s_\pm$ state, it has been suggested that
substantial impurity scattering may induce in-gap states that change the exponential superfluid density to a power-law
dependence, but the exponent is expected to be not smaller than $\sim 2$.\cite{Vorontsov09,Bang09} However, the present
results with exponent significantly smaller than 2 and much closer to 1 as expected for clean superconductors with line
nodes, cannot be explained by these modifications of a full-gap state, but is indicative of well-developed line nodes
in the gap.  We note that our data do not exclude the possibility that some of the bands being fully gapped. Indeed,
the higher temperature behavior of the superfluid density is different to that expected for a single band with line
nodes and instead suggests that some sheets of Fermi surface have a maximum gap below the weak-coupling value, as was
found for MgB$_2$.\cite{Fletcher04}

The fact that the experimental value of $n$ is slightly larger than unity may result from impurity scattering. In the
limit of high levels of disorder, a general gap with line nodes gives $\Delta\lambda(T)\sim T^2$, and the following
formula is often used to interpolate between the clean and dirty limits, $\Delta\lambda(T)\propto
T^2/(T+T^*)$.\cite{Hirschfeld93} The disorder parameter $T^*$ is related to the impurity band width $\gamma_0$. If we use this formula to
fit our data [solid line in Fig.\:\ref{lambda}(b)], we get $T^*=1.3$\,K $\approx 0.04 T_c$, which shows we are close to
the clean nodal limit. A small variable sized upturn in $\Delta\lambda(T)$ is observed at the lowest temperatures
[inset of Fig.\:\ref{lambda}(b)]. This probably originates from amounts (of the order of 0.1\% in volume) of
paramagnetic impurities. This effect is negligible for $T>0.5$\,K ($\sim 0.017T_c$), and is very small in sample \#1,
so this does not affect our conclusion of the existence of low-lying quasiparticle excitations.

%%%%%%%%%%%%Thermal conductivity

\begin{figure}[tb]
%h=here, t=top, b=bottom, p=separate figure pag
\begin{center}%\leavevmode
\includegraphics[width=0.97\linewidth]{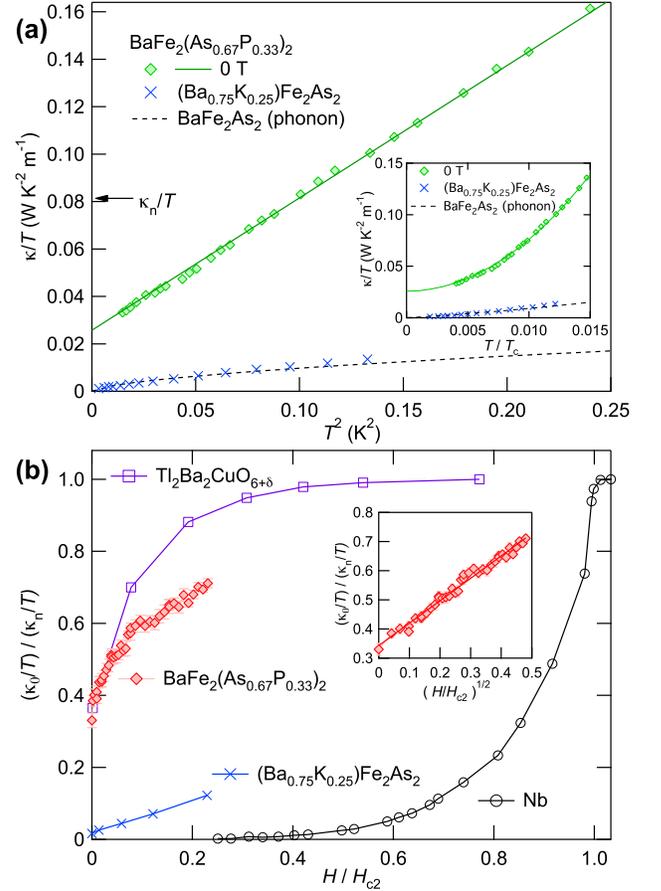}
\caption{ (Color online) (a) Thermal conductivity divided by temperature $\kappa/T$ is plotted against $T^2$ at zero
field (main panel) and on a linear $T$ axis (inset). The solid lines are the fit to the $T^2$ dependence. The reported
data of (Ba$_{0.75}$K$_{0.25}$)Fe$_2$As$_2$ ($T_c\simeq30$\,K) \cite{Luo09} and the phonon contribution estimated in
BaFe$_2$As$_2$ (Ref.\:\onlinecite{Kurita09}) are also shown. (b) Residual $\kappa_0(H)/T$ vs $H/H_{c2}$. Also shown are
the results for Tl$_2$Ba$_2$CuO$_{6+\delta}$ (Ref.\:\onlinecite{Proust02}), (Ba$_{0.75}$K$_{0.25}$)Fe$_2$As$_2$
(Ref.\:\onlinecite{Luo09}), and Nb.\cite{Lowell70} The gradual increase of $\kappa(H)/T$ in
(Ba$_{0.75}$K$_{0.25}$)Fe$_2$As$_2$ has been attributed to the slight modulation of the gap value.\cite{Luo09} Inset
shows the same data in BaFe$_2$(As$_{0.67}$P$_{0.33}$)$_2$ plotted against $(H/H_{c2})^{1/2}$. The line represents the
$\sqrt{H}$ dependence. } \label{kappa}
\end{center}
\end{figure}

Thermal conductivity also provides a probe for the presence of line nodes. First we address the temperature dependence
of $\kappa/T$ in zero field [Fig.\:\ref{kappa}(a)]. In hole-doped (Ba$_{0.75}$K$_{0.25}$)Fe$_2$As$_2$,\cite{Luo09}
$\kappa/T$ is nearly identical to the phonon contribution $\kappa_{ph}(T)$ obtained from non-superconducting
BaFe$_2$As$_2$,\cite{Kurita09} consistent with fully-gapped superconductivity, in which very few quasiparticles are
excited at $T\ll T_c$. In spite of similar values of residual electrical resistivity in the normal
state,\cite{Kasahara09,Luo09} the magnitude of $\kappa/T$ in BaFe$_2$(As$_{0.67}$P$_{0.33}$)$_2$ is strongly enhanced
from that in (Ba$_{0.75}$K$_{0.25}$)Fe$_2$As$_2$. At low temperature the data are well fitted by, $\kappa/T = a T^2
+b$.  The presence of a sizeable residual value $b \simeq25$\,mW/K$^2$m is clearly resolved.

It has been shown that the quasiparticle thermal conductivity in superconductors with sign-changing line nodes is given
by
\begin{equation}
{\kappa}/{T}={\kappa_{0}}/{T} \left(  1+O \left[{T^2}/{\gamma_0^2} \right] \right) \label{kappa_T}
\end{equation}
in the range $k_BT < \gamma_0$, where $\gamma_0$ is the impurity bandwidth.\cite{Graf96} In this case $\kappa_{0}/{T}$
is independent of the impurity content and depends only on the Fermi surface parameters and the slope of the gap near
the nodes.  A rough estimation using parameters for the present material gives
$\kappa_{0}/T\approx22$\,mW/K$^2$m,\cite{estimate} which reasonably coincides with the observed value. The $T^2$-term
in Eq.\:(\ref{kappa_T}) arises from the thermally excited quasiparticles around the nodes.  For
BaFe$_2$(As$_{0.67}$P$_{0.33}$)$_2$ we find that this term is one order of magnitude larger than $\kappa_{ph}/T$.
Recently calculations of $\kappa (T)$ for various candidate gap functions for the Fe-based superconductors have been
reported by Mishra {\it et al.}\cite{Mishra09tcond} For the $s_\pm$ state without sign changing nodes a sizeable
value of $\kappa_{0}/{T}$ is predicted only for very strong pair-breaking scattering (with accompanying strong
reduction in $T_c$) which is incompatible with the low value of $T^*$ found in the $\lambda(T)$ measurements.  For the
case of sign-changing nodes in one or more of the Fermi surface sheets a sizeable $\kappa_{0}/{T}$ is predicted in the
low scattering limit, which is consistent with our experimental results.

%%%%%%%%%%%%%%%%%%%%%%FIG 3%%%%%%%%%%%%%%%
%%%%%%%%%%%%%%%%%%%%%%FIG 3%%%%%%%%%%%%%%%

Next we discuss the field dependence of $\kappa_{0}/T$, which is another independent test of the gap structure. The
most distinguished feature in Fig.\:\ref{kappa}(b) is that $\kappa_0(H)/T$ increases steeply at low fields and attains
nearly 70\% of the normal-state value $\kappa_n/T$ even at 0.2$H_{c2}$ (where $\kappa_n/T \sim 81$\,mW/K$^2$m is
estimated from the Wiedemann-Franz law by using the residual resistivity $\rho_0 \approx30\,\mu\Omega$\,cm and $H_{c2}$
was estimated from heat capacity and torque measurement [inset of Fig.\:\ref{C_T}]). Such a field dependence is
quite similar to that in Tl$_2$Ba$_2$CuO$_{6+\delta}$ with line nodes \cite{Proust02} but is in dramatic contrast to
that in fully gapped superconductors such as Nb.\cite{Lowell70} In fully gapped superconductors, quasiparticles excited
by vortices are localized and unable to transport heat at low fields. In sharp contrast, the heat transport in
superconductors with nodes is dominated by contributions from delocalized quasiparticles outside vortex cores. In the
presence of line nodes where the density of states has a linear energy dependence $N(E)\propto|E|$, $N(H)$ increases
steeply in proportion to $\sqrt{H}$ because of the Doppler shift of the quasiparticle energy.\cite{Volovik} This is
consistent with the field dependence of $\kappa_0(H)/T$ in BaFe$_2$(As$_{0.67}$P$_{0.33}$)$_2$ shown in the inset of
Fig.\:\ref{kappa}(b).

%%%%%%%%%%%Discussion
The present results, (i) the $T$-linear penetration depth, (ii) the large value of $\kappa_{0}/T$ at zero field, (iii)
the $T^2$ dependence of $\kappa/T$, and (iv) the $\sqrt{H}$ field dependence of $\kappa/T$, all indicate that
sign-changing line nodes exist in the gap function of BaFe$_2$(As$_{0.67}$P$_{0.33}$)$_2$. An important question is
then to answer how distinctly different gap structures (with and without nodes) can exist in a BaFe$_2$As$_2$-based
family of Fe pnictides and have comparable transition temperatures. Although the Fermi surface topology is similar in
these two systems,\cite{Kasahara09,Shishido09,Luo08} slight differences in the size and corrugation of hole surfaces
may give rise to the dramatic change of the nodal topology. One possibility is that both systems have a \emph{nodal}
$s$-wave gap function but that the nodes in the electron band are lifted by disorder  in the K-doped system.\cite{Mishra09}  However, as these K- and P-doped samples have very similar normal-state residual resistivities
\cite{Luo09,Kasahara09} as well as similar enhancements in the microwave conductivity below $T_c$,\cite{Tonegawa} this
suggests that they have similar levels of disorder, which makes this scenario unlikely.  According to band-structure
calculations\cite{Kuroki09} the orbital character of one of the hole sheets is very sensitive to the pnictogen height
which changes as As is substituted by P.  This can cause significant changes in the spin-fluctuation spectrum and hence
can change the pairing state.\cite{Kuroki09} This is consistent with the observation of a nodal pairing state in the
Fe-phosphide superconductor LaFePO.\cite{Fletcher09,Hicks09,Yamashita09} However, the nodal gap functions are expected
to give a much lower $T_c$ (as in LaFePO where $T_c \sim 6$\,K)\cite{Kuroki09} so the high $T_c$ of
BaFe$_2$(As$_{1-x}$P$_{x}$)$_2$ remains puzzling.
A recent theory suggested that a competition between orbital and spin fluctuations may lead to line nodes,\cite{Kontani} which needs further investigations.

%%%%%%%%%%%Summary
In summary, from the penetration depth and heat transport measurements, we demonstrate that the high-$T_c$
superconductor BaFe$_2$(As$_{0.67}$P$_{0.33}$)$_2$ has sign-changing line nodes, in sharp contrast to other Fe-based
superconductors which appear to have a fully gapped pairing state. The presence of nodes is strong evidence for a
repulsive pairing interaction such as that provided by antiferromagnetic spin fluctuations.  Understanding the
microscopic origin of these different behaviors remains a challenge for a complete theory of superconductivity in these
materials.

%%%%%%%%%%%Acknowledgments

We thank H. Takeya and K. Hirata for the early-stage collaboration on the crystal growth, and K. Ishida, H. Kontani, K.
Kuroki, and R. Prozorov for helpful discussion. This work is partially supported by KAKENHI, Grant-in-Aid for GCOE
program ``The Next Generation of Physics, Spun from Universality and Emergence'' from MEXT, Japan, and EPSRC in the UK.

%%%%%%%%%%%References


\begin{thebibliography}{99}
\bibitem{estimate}
$\kappa_{0}/T$ is estimated %assuming 4 $c$-axis lines nodes on all Fermi surface sheets 
for a state with line nodes \cite{Graf96} on all Fermi surface sheets by 
$\frac{\kappa_0}{T}\simeq
\frac{\pi\hbar k_B^2}{3e^2\mu_0\lambda(0)^2\Delta_0}$, where $\Delta_0$ is the maximum gap amplitude which we take
equal to $1.7 k_BT_c$.
\end{thebibliography}
\end{document}